# The impact of peer interaction on the responses to clicker questions in an upper-level quantum mechanics course

Ryan Sayer[1], Emily Marshman[2] and Chandralekha Singh[2]
[1]*Department of Physics, Bemidji State University, Bemidji, MN, 56601, USA*
[2]*Department of Physics and Astronomy, University of Pittsburgh, Pittsburgh, PA, 15260, USA*

**Abstract:** In this case study, we investigated the effectiveness of peer interaction on responses to in-class clicker questions in an upper-level quantum mechanics course. We analyzed student performance on clicker questions answered individually and then again after interaction with peers. We also analyzed student performance by topic. In general, the performance on the clicker questions improved after interaction with peers following individual clicker responses. We also find evidence of co-construction of knowledge in that students who did not answer the clicker questions individually were able to answer them correctly after discussion with peers. Finally, we discuss the trends in the percentage of students present in class who responded to the clicker questions in the allotted time as the semester progressed.

## I. INTRODUCTION

Peer collaboration has been used in many formats and instructional settings in physics classes [1-5]. In the approach popularized by Mazur, the instructor poses concrete conceptual problems as multiple-choice clicker questions to students throughout the class. Students first answer the clicker question individually, then discuss the question with their peers and learn by articulating their thought processes and assimilating their thoughts with those of their peers. Instructors also gain valuable feedback about the fraction of the class that has understood the concepts related to the clicker questions at the desired level. The clickers and peer discussions invite active student participation during class time and help them monitor their learning. Peer interaction also provides students an opportunity to be coached by peers who may even be able to discern their difficulties better than the instructor [1-5].

Here, we discuss the findings of a case study in a quantum mechanics (QM) course which involved peer instruction with clickers as part of the in-class instruction. Learning QM is challenging even for advanced students, and prior studies have focused on the difficulties upper-level students have with quantum physics and how to help them learn QM better [6]. We compare students' performance on in-class conceptual clicker questions (concept tests) answered individually after lecture focusing on student difficulties with their performance on clicker questions answered after peer interaction. We also discuss some possible interpretations for the findings. Our case study was designed to shed light on the following research questions: 1) How do students in a QM course perform on clicker questions administered after lectures focusing on their difficulties? 2) Does peer interaction lead to improved performance on clicker questions? 3) What QM concepts do students find challenging as clicker questions? 4) After peer interaction, how often are students in groups able to correctly answer clicker questions if no group member could correctly answer the clicker question before discussion? 5) What are the students' clicker participation patterns during the semester?

## II. METHODOLOGY

This case study was carried out in an upper-level undergraduate elective QM course taught at a large research university. The course consisted of 20 students (mainly physics juniors and seniors) and met on Mondays, Wednesdays and Fridays for 50 minutes. It focused on topics such as the hydrogen atom, identical particles, quantum statistical mechanics, time-independent and time-dependent perturbation theory, and other approximate methods for solving the Time-Independent Schrödinger Equation. In addition to weekly traditional textbook homework problems on the material that was already discussed in the class, students were also assigned weekly pre-lecture readings from the QM textbook by D. Griffiths as homework. They were asked to summarize the assigned reading from the textbook and identify the parts of the material they found challenging. They then submitted their written summaries and feedback on the pre-class reading material they found challenging electronically on the course website before class. The instructor browsed over students' reported difficulties with pre-lecture reading in the summaries they submitted online and tried to tailor the in-class lecture and clicker questions to address students' challenges.

After lectures which focused on student difficulties identified in the pre-lecture reading assignment, students were given multiple-choice *individual concept tests* (ICT) using clickers. Students answered these individually without discussing them with a peer. After answering an ICT, the students were not shown a histogram with the distribution of student responses. Students were then encouraged to discuss the questions in groups of two or three for 1-2 minutes and try to convince their peers sitting next to them about why the response they selected was correct. If students needed more time to discuss the question, additional time was given. After peer interaction, each student individually answered the same clicker question again. We refer to these clicker responses following peer interaction as *group concept tests* (GCT). We observed that students usually discussed the clicker questions with the same one or two peers seated next to them throughout the semester before the GCT. We therefore divided the 20 students into nine groups based on their usual collaborations in the class during clicker questions, which we refer to as groups A through I. We will use these group identifiers to investigate the effectiveness of peer interaction in different groups. After each GCT clicker response, there was a general discussion about each question as a whole class in which the instructor and students participated.

The ICT and GCT questions were developed over a period of more than ten years and went through multiple revisions based on both student and multiple instructors' feedback. Overall, the clicker questions counted as a bonus 2.5% added to the students' total grade. Students were given 80% of the possible points on the ICT and GCT for participating and 100% for answering the questions correctly. Due to time constraints in the classroom, clicker questions were given as both ICT and GCT only during the first six weeks of the course (even during the first six weeks, the instructor did not administer both ICT and GCT for all clicker questions due to time constraint).

Fourteen of the clicker questions for which both ICT and GCT were administered and which are representative of the various QM topics covered in the first six weeks of the course were selected for analysis in this study and will be referred to as *comparison questions*. A list of topics covered by the comparison questions is included in the appendix. The multiple-choice questions were first graded as correct or incorrect to determine the "unadjusted" scores. The scores were then adjusted using an established procedure to account for the possibility of guessing, which we refer to as "adjusted" scores [7]. We show in the results section that while the quantitative features of our findings depend on whether the scores are unadjusted or adjusted, the qualitative features are similar in both cases.

## III. RESULTS AND DISCUSSION

The average scores on the ICT and GCT for all 14 comparison questions were averaged over all students, as shown in Table I. Overall, there was an improvement in students' performance from ICT to GCT, and the difference between the means of ICT and GCT was significant.

**Performance by student averaged over all comparison questions**: Figure 1 shows that, on average, students showed significant improvement from the ICT to GCT after discussing the questions with their peers. The same qualitative trend is observed for both the unadjusted and adjusted scores. Figure 1 shows that, with the exception of a few outliers, all students performed well on GCT regardless of their performance on ICT. In many of the groups, both group members showed improvement after discussing the questions (e.g., group D). Also, students who initially underperformed on clicker questions often benefited the most from interactions with their peers, as measured by the difference in the average performance on GCT vs. ICT. For example, in Group A, one student performed much better on ICT questions than the other, but both students performed well on the GCT after discussion with each other. However, Fig. 1 also suggests that sometimes the peer interaction did not appear to help certain students (e.g., one of the students in Group F). Consideration of the overall class grades of students in groups in which one peer did not benefit as much does not suggest any obvious pattern or academic reasons for the lack of benefit of interaction for all peers.

**TABLE I**. Unadjusted and adjusted student percentages (averaged over all students and all comparison questions) on the ICT and GCT, with *p*-values for comparison.

|  | ICT | GCT | *p* |
|---|---|---|---|
| Unadjusted % | 69 | 85 | 0.003 |
| Adjusted % | 48 | 73 | 0.009 |

Several factors foster productive group discussions. Interaction with peers provides opportunity for clarifying difficulties, especially if there are diverse opinions. Also, students who have recently learned the concepts understand other students' difficulties better than the instructor and may be in a better position to help their peers if they are comfortable discussing their thought processes with their peers. In supportive environments, peer interaction generally helps all students since discussing and articulating concepts gives further clarity to thought processes and can help all students develop a better grasp of physics concepts [1-3]. Peer interaction keeps students alert and on their toes because they must explain their reasoning to peers.

For students who benefited significantly from peer interaction, struggling to answer an ICT before discussing their thought processes with their peers for that question may have been productive and helped them focus on the discussions with their peers [8]. Another reason why peer interaction may have helped students perform better on the GCT is that the peer interaction was extended over a period of time and students may have begun to realize that their peers struggled with the same concepts. They may then attribute their struggles to the difficulty of the subject matter rather than personal factors. This type of supportive class dynamics has the potential to make students less anxious while learning the QM concepts.

**Performance by question averaged over all students**: We next compare the average unadjusted and adjusted performances for each comparison question on the GCT versus the ICT, as shown in Fig. 2. Each data point in Fig. 2 represents the average score of all students on a particular question. The students on average showed improvement for most questions after peer interaction, and the trends were similar for both the unadjusted and adjusted scores. Student performance reached the ceiling on the GCT for certain questions involving simple application of principles, such as Question II which concerns Hund's rule for total orbital angular momentum. However, Fig. 2 shows that, on average, students did not improve from the ICT to the GCT on Question III, which asked them to determine the probability of finding an electron in a hydrogen atom at a position between $r$ and $r + dr$ from the nucleus. Question III involves synthesis of mathematical knowledge and skills with knowledge of quantum physics. It may be advantageous to first break down such multiple-choice synthesis problems into separate multiple-choice sub-problems (to be posed as ICT and GCT) to make them more manageable for students to think about and discuss with peers before combining them into one clicker question.

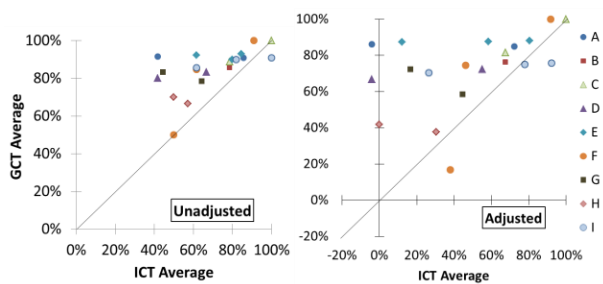

**FIG 1**. Each student's unadjusted (left) and adjusted (right) performance on GCT vs. ICT, averaged across all comparison questions. The color/symbol indicates the group to which that student belonged for the GCT.

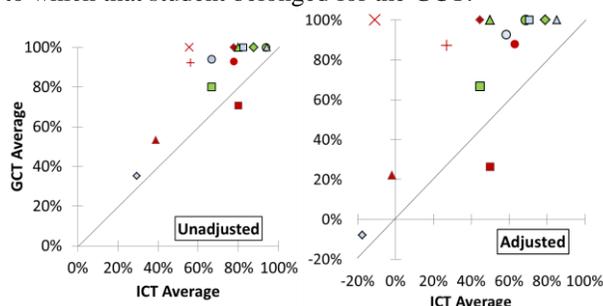

**FIG 2**. Unadjusted (left) and adjusted (right) performance on GCT vs. ICT for each comparison question, averaged across all students. The color/symbol indicates the question topic.

**Co-construction of Knowledge**: Prior research suggests that, even with minimal guidance from the instructors, students can benefit from peer interaction [3]. In particular, those who worked with peers not only outperformed an equivalent group of students who worked alone on the same task, but collaboration with a peer led to "co-construction" of knowledge in 29% of the possible cases in that study [3]. Co-construction of knowledge occurs when neither student who engaged in the peer collaboration was able to answer the questions before the collaboration, but both were able to answer them after working together. In order to determine whether peer interaction shows evidence of co-construction in this case study, we analyzed performance of students on GCT depending upon the ICT performance of the peers in each group for all questions. Row 1 (with data) in Table II represents the situation in which all group members answered an ICT incorrectly and shows the percentages of all clicker questions for which all group members answered the corresponding GCT incorrectly (column 1 with data), one group member answered incorrectly (column 2 with data), and all group members answered correctly (column 3 with data). For example, Row 1 (with data) in Table II shows that when all group members answered an ICT incorrectly they all answered the corresponding GCT correctly (i.e., they "co-constructed" knowledge) 31% of the time. Row 2 (with data) in Table II shows that when only one group member answered an ICT correctly, all group members answered a GCT correctly 77% of the time. Row 3 (with data) shows that when all group members answered an ICT correctly, all of them answered the GCT correctly 98% of the time.

**TABLE II.** Percentage of GCT clicker questions for which (1) both group members answered incorrectly, (2) one member answered correctly and one incorrectly, and (3) both answered correctly, as a percentage of ICT responses in each category.

|  |  | GCT | | | |
|---|---|---|---|---|---|
|  |  | (1) | (2) | (3) | Total |
| ICT | (1) | 61% | 8% | 31% | 100% |
|  | (2) | 19% | 4% | 77% | 100% |
|  | (3) | 2% | 0% | 98% | 100% |

**Clicker trends over time**: We also compared students' average gains from the ICT to the GCT for each of the first six weeks of class instruction, as shown in Fig. 3. We hypothesized that student groups may become more cohesive and their discussions more productive as the semester progressed, resulting in larger gains from the ICT to the GCT. Figure 3 shows that for the first five weeks of the course, the students had larger gains from the ICT to the GCT each week than they had in the previous week. This suggests that there may be a "learning curve" to peer instruction in groups, and that students may need some time to familiarize themselves with the communication styles and discussion approaches of peers (in addition to the style of the instructor). One possible reason for the dip in Fig. 3 in week 6 may be the difficulty associated with the challenging concept of degenerate perturbation theory.

Figure 4 shows the average non-response percentage for the whole class for each week of instruction. The first two weeks of the course had much higher non-response rates on both the ICT and the GCT. However, the non-response rates declined greatly after the first two weeks of the course and stayed low for the rest of the course. A missed response to a clicker question is only counted as a non-response if the student was present in the classroom when the clicker question was given. Attendance in the class was generally greater than 80%. One interpretation of Figs. 3 and 4 is that students needed time to familiarize themselves with clicker question procedures and with their peers and develop the habit of regularly clicking in response to all clicker questions posed. The higher non-response rates on the GCT compared to the ICT could partly be due to students disagreeing with their peers about their responses, getting distracted by discussions and not clicking despite the instructor's reminder before time was up.

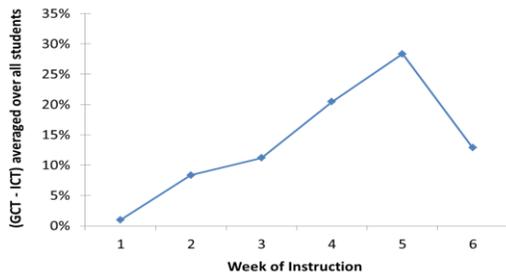

**FIG 3**. (GCT - ICT) scores for each week of instruction (averaged over all students and all questions for that week).

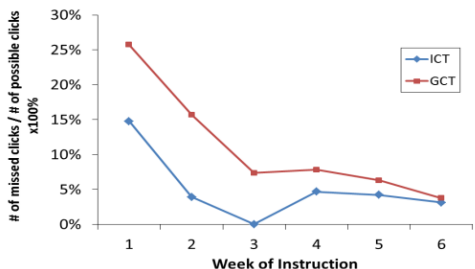

**FIG 4**. Student non-response on ICT (blue) and GCT (red) as a percentage of total possible responses per week of instruction.

## IV. SUMMARY

In general, students' performance improved from the ICT to the GCT regardless of the initial difficulty of the clicker questions in individual administration. Students who scored below average on the ICT showed greater improvement in performance on the GCT after peer interaction. Students were able to "co-construct" knowledge in a peer interaction so that all members of the group selected the correct response on the GCT for 31% of the clicker questions for which all group members responded incorrectly on the ICT. For each of the first five weeks of the course, student performance improved more, on average, from the ICT to the GCT than it did in the previous week. Also, non-response rates on the in-class clicker questions started at or above 15% at the beginning of the semester but decreased in later weeks of the course. One possible reason for this trend is that the students may need a few weeks to familiarize themselves with the in-class clicker procedures and group work. We also find that for a given student, the cumulative non-response rates for the entire semester was generally higher on the GCT than on the ICT. To the best of our knowledge, such trends in the clicker responses have not been reported in prior research in physics classes.

## ACKNOWLEDGEMENTS

We are grateful to the National Science Foundation for award PHY-1505460.

## APPENDIX

This is a list of the topics covered by comparison questions:
- I-II. Hund's rules for the total spin (S) and total orbital angular momentum (L).
- III. Probability of finding an electron between a distance $r$ and $r + dr$ from the nucleus of a hydrogen atom.
- IV-V. Spin configuration of electrons for a helium atom in the ground state and in an excited state.
- VI-VII. Fermi energy and total energy associated with valence electrons of copper cubes of different sizes at temperature $T = 0K$.
- VIII. Change in total energy associated with valence electrons as the volume of a copper cube is changed but the number of atoms is kept fixed.
- IX-X. Non-interacting distinguishable particles and bosons in a one-dimensional infinite square well.
- XI. Non-interacting fermions in single particle states.
- XII. Given that the perturbing Hamiltonian $\hat{H}'$ and the unperturbed Hamiltonian $\hat{H}^o$ both commute with some Hermitian operator $\hat{A}$, do they necessarily commute with each other?
- XIII-XIV. Is an eigenstate $|a\rangle/|c\rangle$ corresponding to a degenerate/non-degenerate subspace of $\hat{H}^o$ necessarily a "good" state for a given perturbing Hamiltonian $\hat{H}'$?

[1] E. Mazur, Peer Instruction, Prentice Hall, N.J., (1997).
[2] P. Heller et al., Am. J. Phys. **60**, 627 (1992).
[3] Singh, Am. J. Phys. **73**, 446 (2005); arXiv:physics/0207106
[4] A. Mason and C. Singh, Am. J. Phys. **78**, 748 (2010); AIP Conf. Proc., Melville New York **1289**, 41 (2010); Mason and Singh, The Physics Teacher **54**, 295 (2016).
[5] K. Miller et al., Phys. Rev. ST PER **11**, 010104 (2015).
[6] C. Singh, Am. J. Phys. **69**, 885 (2001); **76**, 277 (2008); **76**, 400 (2008); M. Wittmann et al., Am. J. Phys. **70**, 218 (2002); D. Zollman et al., Am. J. Phys. **70**, 252 (2002); G. Zhu and C. Singh, Am. J. Phys. **79**, 499 (2011); **80**, 252 (2012); Phys. Rev. ST PER **8**, 010117 (2012); **8**, 010118 (2012); **9**, 010101 (2013); E. Marshman and C. Singh, Phys. Rev. ST PER **11**, 020119 (2015); Eur. J. Phys. **37**, 024001 (2016); arXiv:1602.05720; arXiv:1602.05461; arXiv:1602.06374; C. Singh and E. Marshman, Phys. Rev. ST PER **11**, 020117 (2015); B. Brown et al., Phys. Rev. PER, **12**, 010121, (2016); G. Passante et al., Phys. Rev. PER **11**, 020111 (2015); arXiv:1602.05655; arXiv:1602.05664; arXiv:1509.07740; arXiv:1603.02948; arXiv:1603.06025.
[7] C. Bernaards and K. Sijtsma, Multivariate Behavior Res. **34**, 315 (1999).
[8] D. Schwartz and J. Bransford, Cog. and Inst. **16**, 475 (1998); M. Kapur, Cog. and Inst. **26**, 379 (2008).